# Control System Design for Tricopter using Filters and PID controller

**P Venkata Vishal**

M-Tech (Embedded systems technology), SRM
UNIVERSITY, CHENNAI,
603203, INDIA,
Email: venkatavishal_partha@srmuniv.edu.in

**V. Natarajan**

Professor, Department of Electronics and
Communication Engineering,
SRM UNIVERSITY, CHENNAI-603203, INDIA

*Abstract*—The purpose of this paper is to present the control system design of Tricopter. We have presented the implementation of control system in software in this paper. Tricopter's control system mainly consists of two parts: Complimentary filter and PID block. The angles along X, Y and Z axis are taken from the complementary filter which acts as a feedback block. We have used the combination of gyroscope and accelerometer for finding the angle. We have also shown the role of the complimentary filter in finding out the angle along X, Y, and Z axis instead of using gyroscope and accelerometer directly. The second main part is the PID Controller which calculates the error in angle along X, Y and Z axis and produces an output signal which reduces error. We have shown the importance of the constant parameters of PID Controller. The results of this paper are tested on an actual Tricopter and also plotted in the form of graph using Matlab and Processing software.

## I. INTRODUCTION

Tricopter is basically a UAV (Unmanned Aerial Vehicle) with three rotors, two of them are fixed rotors and the third rotor is attached to a servo motor. We call the third rotor as the tail rotor. The tail rotor twists around X axis. This mechanism allows to counter act the anti-torque generated by the Tricopter. All the three rotors are used to lift and propel the Tricopter. The Tricopter uses three fixed pitched propellers out of which two rotors rotate in anticlockwise (ACW) and the other rotor in clockwise direction (CW).

Initially the UAV's developed were not stable due to the several factors. Few of the major factors were unavailability of powerful and compact electrical motors, ESC (Electronic Speed Controller) for controlling the electrical motor. Gyroscopes and accelerometers were mechanical devices and they were very difficult to interface with electronic devices. High speed and compact microcontrollers did not exist which is necessary to process Algorithms for balancing UAV's. Due to rapid growth and progression in the field of VLSI (Very large Scale Integration), power electronics and MEMS (Micro Electromechanical System) have significantly reduced the size of sensors, microcontrollers and improved its throughput and performance. BLDC motors (Brushless DC) developed recently, has a very high power to weight ratio and needs very less care and maintenance. The BLDC motors are normally three phased, so supply of DC power directly will not run the motors. It requires a special device called Electronic Speed Controllers (ESC) which generates three phase ac signal constantly to keep the motor running. Thanks to these advancements, it has becomes easy to write algorithms for microcontrollers to control the speed of all rotors to accomplish steady flight. The ESC is a controller board for controlling the speed of BLDC motors. It has a battery on the input side and a three phase output for running the BLDC motor. Each ESC is controlled independently by PWM (Pulse Width Modulated) signal. Another input to the ESC is a PWM (Pulse Width Modulated) signal [8] that controls each ESC independently.

## II. SYSTEM DESIGN

There are basically three parts in our system: Input section, Output section and PID Section. The Throttle signal given to microcontroller is converted into PWM Signal and is given to the ESC which controls the speed of motor. The PWM signal is obtained by using hardware timers of the microcontroller. $X_{roll}$, $Y_{pitch}$ and $Z_{yaw}$ are set point angles along X, Y and Z respectively. Current angle is given by complimentary filter using which the error is calculated and given to PID Controller. The PID generates PWM signal and is added or subtracted from the throttle's PWM signal to maintain the set point [2].

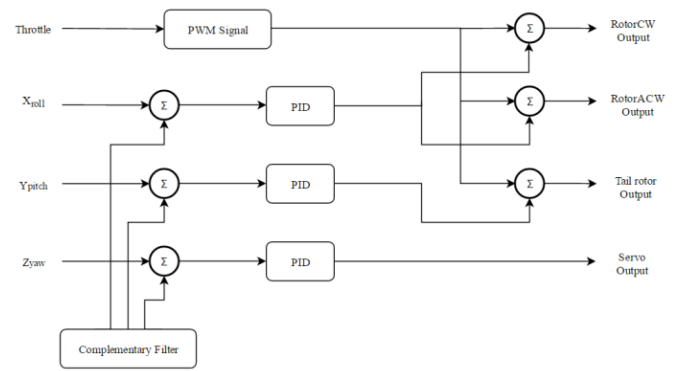

Fig. 1. System Design

## III. Sensor Integration

To find the position of Tricopter we have used IMU (Inertial Measurement Unit) MPU6050 sensor which gives us the value along X axis Y axis and Z axis. IMU consist of 3 axis gyroscope and 3 axis accelerometer. It contains ADC which is 16 bit for digitizing the output of gyroscope and accelerometer. The data is transferred using an I2C bus to the microcontroller from MPU6050 at 400KHz. SPI bus can also be used to transfer data at 1MHz [1].

Accelerometer measures magnitude of gravity that is force/mass. It can be used as an inclinometer in Tricopter. If Tricopter lifts up then the accelerometer reads a positive value. If Tricopter goes down then the accelerometer reads a negative value. It cannot tell the difference between stationary incline changes versus short term transitional acceleration [8]. So an accelerometer alone cannot be used as a feedback in Tricopter. We have presented the accelerometer's performance in Fig. 3.

Gyroscope measures rotational motion. It senses changes in the orientation. When stationary it outputs zero. When it is spinning it outputs a positive or negative value. The mechanical vibrations generated by the turbulence of the motors can badly affect the quality of the output from the gyroscope. Therefore gyroscope alone cannot be used for the feedback in Tricopter [8]. We have presented the Gyroscope's performance in Fig. 4.

## IV. Filter Design

Accelerometer measures static angles i.e. fast changes are not registered. Gyroscope measure changes in the orientation i.e. slow changes are not registered. So to minimize these problems we have used complementary filters, which is a combination of LPF (Low Pass Filter) and HPF (High pass Filter) [4].

## V. Complementary Filter

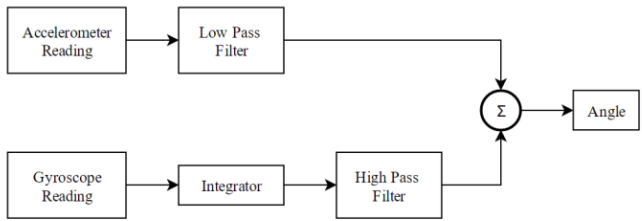

Fig. 2. Complementary filter

Integrator: The need of integrator is to take the sum of finite number of samples of gyroscope output taken at constant time interval. It can be implemented in software as:

Angle + = gyro * dt

LPF: We have used the LPF to consider long term changes and discard short term fluctuations. The output from the gyroscope is given to the LPF as shown in the Fig. The output of the LPF is implemented using C language as shown in the equation below

Angle += (0.98) * Angle + (0.02) * x_acc;
The term (0.02) * x_acc acts as a LPF.

HPF: We have used HPF to consider short duration changes and discard signals that are steady over time. It can be used to cancel out the drift in Tricopter.

Sample period: it is the time that takes to complete one program loop. If the sample rate is chosen to be 100 Hz, the sample period will be 0.01 s [2-5].

Final equation of the complementary filter we have used in software implementation using C language is
X_angleC = 0.93 * (X_angleC + newGyroRate * dtC) + 0.07 * newAccAngle;

X_angleC is a variable of type float. This is the angle that has to be returned from the function. newGyroRate is a variable of type float, it is the angle measured using the gyroscope. dtC is a variable of type float, it is the sample time for the gyroscope. newAccAngle is a variable of type float, it is the angle measured using the accelerometer.

For our Tricopter experimental results are shown in the Fig. 3, Fig. 4 and Fig. 5. It is the response of the accelerometer, gyroscope and complementary at 180 degree respectively [3-4].

Table I Experimental readings of complementary filter

| Angle (deg) | Output of accelerometer X (roll) | Output of gyroscope X (roll) | Output of complementary filter X (roll) |
|---|---|---|---|
| 30 | 35.34 | -22.58 | 31.24 |
| 60 | 63.34 | -328.11 | 61.57 |
| 90 | 94.20 | -325.31 | 91.05 |
| 120 | 117.32 | -53.15 | 122.05 |
| 160 | 165.75 | -676.27 | 161.24 |

Table II Experimental readings of complementary filter

| Angle (deg) | Output of accelerometer Y (pitch) | Output of gyroscope Y (pitch) | Output of complementary filter Y (pitch) |
|---|---|---|---|
| 30 | 33.54 | -119.05 | 31.85 |
| 60 | 63.52 | -141.68 | 61.69 |
| 90 | 93.35 | -271.36 | 91.56 |
| 120 | 124.52 | -295.65 | 122.55 |
| 160 | 157.85 | -320.14 | 153.65 |

Table III Experimental readings of complementary filter

| Angle (deg) | Output of accelerometer Z (yaw) | Output of gyroscope Z (yaw) | Output of complementary filter Z (yaw) |
|---|---|---|---|
| 30 | 31.84 | -21.67 | 32.32 |
| 60 | 62.65 | -326.91 | 62.65 |
| 90 | 92.88 | -323.46 | 92.86 |
| 120 | 122.72 | -51.32 | 123.85 |
| 160 | 163.79 | -674.35 | 164.81 |

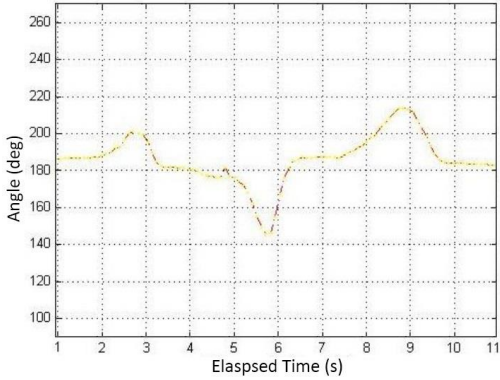

Fig. 3. Response of accelerometer

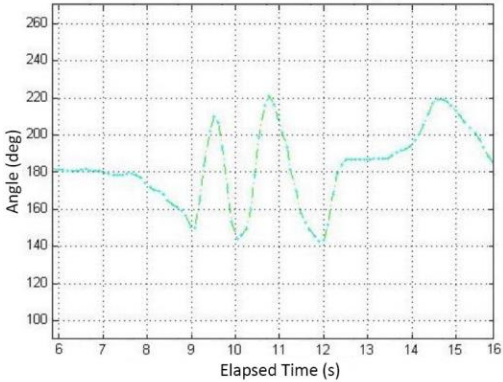

Fig. 4. Response of gyroscope

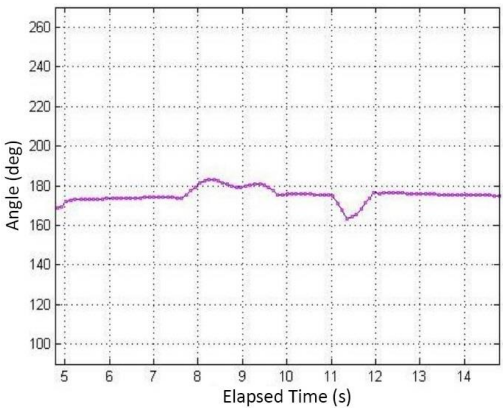

Fig. 5. Response of complementary filer

## VI. PID CONTROLLER

The PID stands for Proportional, Integral and Derivative controller. In general the control systems can be briefly classified into two types: Open loop control system and closed loop control system. In open loop control systems the output is computed based on the set point where as in the closed loop control system it uses feedback to compute the output. Closed loop control systems have the ability to adapt to the changing conditions and aid them to make the unstable process stable. Here we use a closed loop control system design. The control loop feedback is implemented using a PID loop. The difference between the measured value and the set point is the error signal which is given to the PID block. The PID controller tries to minimize the error in the outputs by modifying the process control inputs.

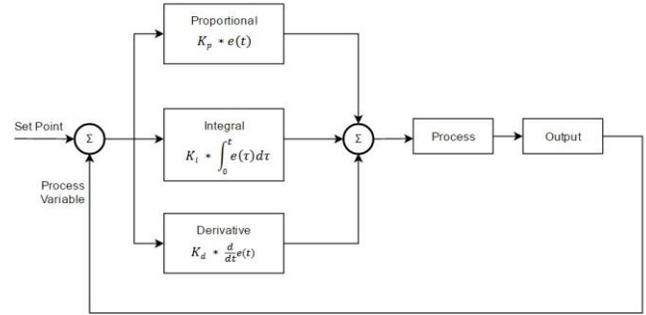

Fig. 6. PID controller

The PID controller has three constant parameters they are proportional, integrative and derivative. P relies on the present error, I on the accumulation of past errors and D predicts the future errors, based on the current rate of change. The proportional controller ($K_p$) will have an influence in reducing the rise time, but never eliminates the steady state error. The proportional block depends only on the difference between set point and process variable. This difference is called the error term. The job an integral block ($K_i$) in the PID controller is to eliminate the steady state error for a constant input, but this block may make the transient response slower. This block sums the errors over time; hence this term continually increases over time. The derivative block ($K_d$) in the PID controller makes the system more stable by reducing the overshoot and improving the transient response [7].

Proportional term: $P_{out} = K_p * e\,(t)$

Integral term: $I_{out} = K_i * \int_0^t e(\tau) * d\tau$

Derivative term: $D_{out} = K_d * \frac{d}{dt} e(t)$

So to achieve stable flight of Tricopter we need to find the appropriate values of $K_p$, $K_i$ and $K_d$. These coefficients individually have different effects on Tricopter flight.

Proportional gain coefficient: $K_p$ coefficient of proportional block is responsible for sensitivity and reactivity for angular change. This coefficient tells how fast it has to react to the angular change. If it is too low the Tricopter tends to be very sluggish and unsteady. If it is too high then the Tricopter starts to oscillate [7].

Integral gain coefficient: $K_i$ coefficient can increase the accuracy of the angular position. For example when the Tricopter is disturbed and its angle changes 10 degrees, in theory it knows how much the angle has changed and will return 10 degrees. Without having this term, the opposition does not last as long. However, when this coefficient gets too high the Tricopter begins to have slow reaction and it decreases the effect of the Proportional gain as consequence, it will start to oscillate just like having high P gain, but at a lower frequency [7].

Derivative gain coefficient: $K_d$ coefficient lets the Tricopter to reach more rapidly the desired attitude. In practice it will increase the reaction speed. D gain makes the Tricopter more sensitive. Higher the $K_d$ coefficient lesser the oscillations and shortens the settling time [7].

Final Values of $K_p$, $K_i$ and $K_d$ tested in the Tricopter is shown in the Table IV

Table IV Empirical values of $K_p$, $K_i$, $K_d$

| $K_p$ | $K_i$ | $K_d$ |
|-------|-------|-------|
| 1.41  | 0.91  | 1.31  |

## VII. RESULTS

The results are shown in the Fig. 7, Fig. 8 and Fig. 9, these figures show the PID graphs for pitch, roll and yaw of Tricopter. In the graph green line shows the set point of PID. Red line shows the offset and blue line shows the output of PID [5-8].

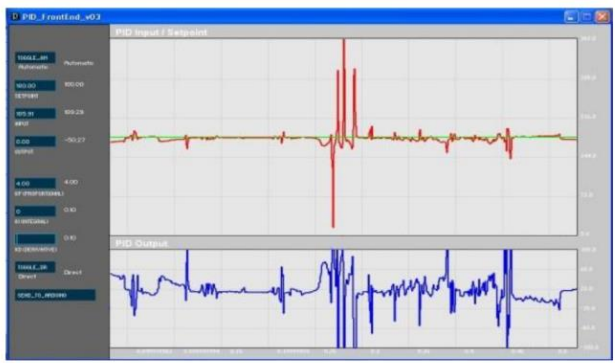

Fig. 7. Response of pitch

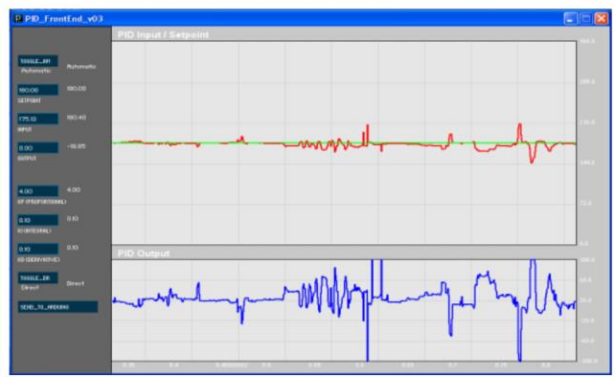

Fig. 8. Response of roll

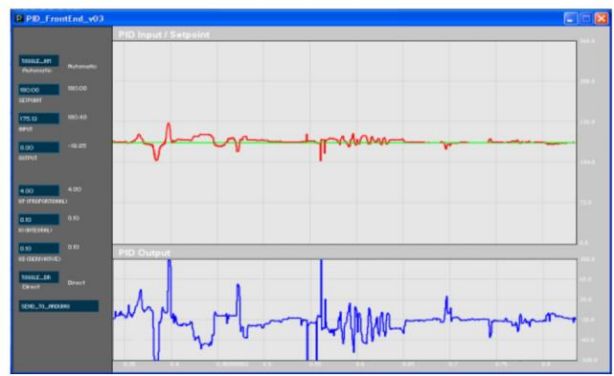

Fig. 9. Response of yaw

## VIII. CONCLUSION

In this paper we have presented the control system design for a Tricopter in view of software implementation. We have got the result from the output of the complementary filter by computing the input from accelerometer and gyroscope. The response of PID controller is acceptable. From the calculations of angle we get the values and the graphs.

### ACKNOWLEDGMENT

The authors would like to express their earnest gratitude to Department of Electronics and Communication Engineering of SRM University, Chennai, India for providing us the required facilities and undeterred support throughout the duration of the project work. The Author P Venkata Vishal would also like to thank his family and friends for their support during the course of the work.